\documentclass[12pt,letter,logo,copyright,nonumbering]{xds}
\usepackage{subcaption}
\usepackage{dblfloatfix}
\usepackage{ulem}
\usepackage{float}
\usepackage{caption}
\usepackage{dramatist}
\usepackage{xspace}
\usepackage{pifont} 
\usepackage{multirow}
\usepackage{tcolorbox}
\usepackage{xltabular}
\usepackage{longtable}
\usepackage{mdframed}
\usepackage{hyperref}
\usepackage{tocloft}
\usepackage{amsfonts}
\usepackage{amsmath}
\usepackage{amssymb}
\usepackage{lineno}
\usepackage{multirow}
\usepackage{adjustbox}

\usepackage[bottom]{footmisc}

\usepackage{CJKutf8}
\usepackage{setspace}

\usepackage{graphicx}

\usepackage{dsfont}
\usepackage{array} 
\usepackage{tabularx} 
\usepackage{xcolor} 

\usepackage{lipsum}  
\usepackage{multicol} 

\makeatletter
\def\@BTrule[#1]{%
  \ifx\longtable\undefined
    \let\@BTswitch\@BTnormal
  \else\ifx\hline\LT@hline
    \nobreak
    \let\@BTswitch\@BLTrule
  \else
     \let\@BTswitch\@BTnormal
  \fi\fi
  \global\@thisrulewidth=#1\relax
  \ifnum\@thisruleclass=\tw@\vskip\@aboverulesep\else
  \ifnum\@lastruleclass=\z@\vskip\@aboverulesep\else
  \ifnum\@lastruleclass=\@ne\vskip\doublerulesep\fi\fi\fi
  \@BTswitch}
\makeatother

\addto\extrasenglish{
}

 {\begin{list}{}%
         {\setlength{\leftmargin}{#1}}%
         \item[]%
 }
 {\end{list}}

\reportnumber{001} 
\renewcommand{\today}{}

\usepackage[english]{babel}
\usepackage{microtype}
\usepackage{graphicx}

\usepackage{booktabs} 
\usepackage{xcolor}
\usepackage{listings}

\usepackage{tabularx}

\usepackage{hyperref}

\usepackage[square,numbers]{natbib}
\renewcommand{\today}{}

\newcommand{\sysname}{ReviveMoE}
\newcommand{\AtoE}{\lstinline|A2E|}
\newcommand{\EtoA}{\lstinline|E2A|}
\newcommand{\dispatch}{\lstinline|dispatch|}
\newcommand{\combine}{\lstinline|combine|}

\def\Snospace~{\S{}}
\addto\captionsenglish{%
}
\addto\extrasenglish{%
}

\usepackage{subcaption}

\usepackage{tikz}
\DeclareRobustCommand{\circled}[1]{\tikz[baseline=(char.base)]{
            \node[shape=circle,draw,inner sep=0.5pt] (char) {\small #1};}}

\title{\sysname: Fast Recovery for Hardware Failures in Large-Scale MoE LLM Inference Deployments}

\author[1]{Haley Li}
\author[1,3]{Xinglu Wang}
\author[2]{Cong Feng}
\author[2]{Chunxu Zuo}
\author[1]{Yanan Wang}
\author[1]{Hei Lo}
\author[1]{Yufei Cui}
\author[2]{Bingji Wang}
\author[2]{Duo Cui}
\author[2]{Shuming Jing}
\author[2]{Yizhou Shan}
\author[1]{Ying Xiong}
\author[3]{Jiannan Wang}
\author[1]{Yong Zhang}
\author[1]{Zhenan Fan}
\affil[1]{Huawei Technologies Canada}
\affil[2]{Huawei Technologies China}
\affil[3]{Simon Fraser University}

\begin{document}

\maketitle
\begin{abstract}
As LLM deployments scale over more hardware, the probability of a single failure in a system increases significantly, and cloud operators must consider robust countermeasures to handle these inevitable failures. A common recovery approach is to simply restart the LLM serving instance; however, this is costly in model-as-a-service (MaaS) inference settings, where reloading model weights and recompiling computation graphs can introduce significant delays to incoming requests. We propose \sysname, a method for rapid failure recovery in large-scale LLM deployments without restarting the serving instance. \sysname\ is designed to support both the traditional LLM architecture, which collocates MoE and attention on the same hardware, and the disaggregated architectures, which separate MoE from attention. Integrated into Huawei Cloud's MaaS, \sysname\ is built on top of Huawei's xDeepServe serving platform and the XCCL communications library.
\end{abstract}
\abscontent
\newpage
\tableofcontents
\newpage
\section{Introduction}
As large language models (LLMs) grow in terms of parameter counts, their hardware requirements grow accordingly. For example, the 671B parameter DeepSeek V3~\cite{dsv32025} required at least 320 GPUs for its decoding stage deployment. However, larger cluster sizes increase the likelihood of hardware failures~\cite{kokolis2025}. Prior works claimed rates of 1 failure per 7.9 hours on 1024 GPU clusters~\cite{kokolis2025} and 2-5\% of nodes experience hardware failures each month~\cite{gershon2025}. Even if hardware failures are not cluster-wide, a single hardware failure can still halt the entire system, especially in communications-reliant mixture of experts (MoE) models like DeepSeek. As hardware demands grow, large-scale deployments motivate the need for fault tolerant inference designs to mitigate inevitable failures.

Current research on LLM reliability is focused on failure prevention~\cite{kokolis2025} or training-time recovery~\cite{mohan2021, duan2024, gupta2024, zhang2025}. While prevention is important, failures are inevitable at large hardware scales, and countermeasures are necessary. Training recovery is not well suited for the inference setting, as it focuses on checkpointing to avoid wasted training steps. 
Unlike training, inference does not need to perform checkpointing, as the weights remain static throughout the inference lifecycle. Additionally, due to inference's customer-facing nature, faster recovery is needed to meet service level objectives (SLOs). This leaves a gap where there is little research in fault recovery for LLM inference. 

At Huawei Cloud, we are interested in serving large language models (LLMs) on our CloudMatrix384 infrastructure~\cite{zuo2025} using our in-house xDeepServe \cite{xiao2025} serving platform. Production deployments of DeepSeek~\cite{dsv32025}, Kimi~\cite{kimik22025}, and Qwen~\cite{qwen32025} span hundreds of Ascend neural processing units (NPUs), where failures are unavoidable. Failures can be addressed by restarting the instance, but this is costly in model-as-a-service (MaaS) settings, as reinitializing the instance can degrade the service for several minutes, leading to SLO violations.

\begin{figure}[bpht]
    \centering
    \includegraphics[width=0.5\linewidth]{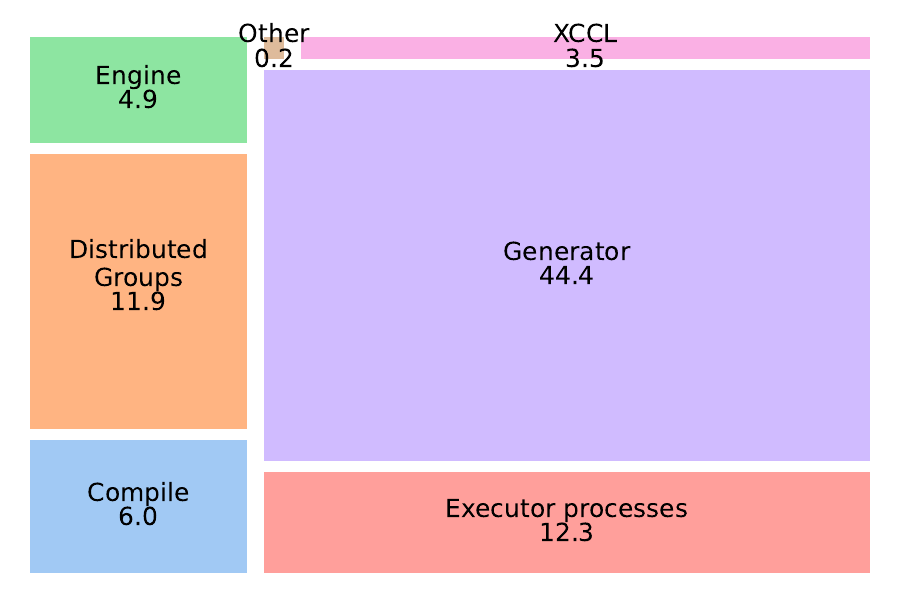}
    \caption{Breakdown of the time taken in seconds for a cached reinitialization of a DeepSeek V3 instance on 80 NPUs (total 83.1 s). ``Compile'' refers to a cached compile.}
    \label{fig:motivation_reinit}
\end{figure}

A full reinitialization resets the system but performs many unnecessary steps. To identify such necessary procedures, we break down the major components of an instance reinitialization in \autoref{fig:motivation_reinit}. 
The largest component, the generator, involves model instantiation and weight loading, which are often unnecessary after failure. Redundancy mechanisms such as data parallelism (DP) and redundant experts~\cite{dsv32025} preserve model weights on surviving hardware, allowing us to oftentimes skip costly weight loading, thus eliminating the largest time cost. Many CPU processes, such as the engine and executors, do not need reinitialization; only failed or hanging processes must be terminated.
Therefore, effective recovery should focus on isolating failed processes, restoring communications, and re-establishing computation graphs, all while maintaining user requests, model weight integrity (in cases where replication is not sufficient), and scheduler state.

We present \sysname, our design for a failure recovery system for xDeepServe. We fit \sysname\ into two deployment designs. The first is the traditional LLM deployment design that collocates the MoE and attention modules on the same hardware. The second is the disaggregated case that separates MoE and attention onto separate NPUs. We propose the following contributions:

\begin{itemize}
    \item \textbf{Failure isolation and state recovery.} \sysname\ confines faults to failed hardware and migrates active sequences to healthy ranks while preserving prompts and decoded tokens. A log-based recovery mechanism restores the block table to its pre-failure state.
    \item \textbf{Weight integrity.} Leveraging DP and redundant experts, \sysname\ avoids unnecessary weight reloads. When redundancy is limited, it supports (1) role switching redundant DP ranks to MoE roles, or (2) tolerating expert loss with minor accuracy degradation. We show the accuracy loss is negligible in large (EP32) deployments (\autoref{sec:eval:accuracy}).
    \item \textbf{Communications and graph compilation.} \sysname\ reconstructs communication domains using new rank assignments to handle failed hardware. This procedure requires recompiling the computation graph before inference can continue. We split compilation into precompile and cached compilation stages. Compared to full graph recompilation from scratch (up to 15 minutes for DeepSeek V3), our approach leverages precompiled graphs for common failure scenarios in the cached compilation stages, reducing recompilation time to under 10 seconds.
\end{itemize}

In deploying \sysname, we reduce recovery time from 83.1 seconds for a complete instance restart to 10.2 seconds (87.8\% reduction). Even in the worst case, which requires a role switch and weight loading from disk, \sysname\ still reduces recovery time by 36.6\%.

\section{Background}

\subsection{Mixture of Experts Models}
Traditional transformer architectures encounter many practical issues with training and computational cost when scaling to large parameter counts~\cite{lepikhin2021, dai2024}.  Mixture of experts (MoE) models combat these issues by introducing sparsity. The first component involves replacing the dense feedforward networks (FFN) after the attention blocks with an MoE layer, a set of experts, each acting as an FFN. A gating function is then inserted after each attention module. It computes a score over the experts for each token, and tokens are routed to their top-k experts. This only activates a subset of the MoE layer for each token, thus achieving sparsity. 

To support MoE models in distributed environments, they are often deployed with a combination of data parallel (DP), tensor parallel (TP), and expert parallel (EP) configurations. DP replicates the attention modules over the system, so that each DP instance contains the full set of attention weights and can process batches independently from other DP instances. TP shards the weights of the attention module and any dense feedforward layers over multiple NPUs, and collective communications operations are required to dispatch and combine their inputs and outputs. EP~\cite{lepikhin2021} divides the set of experts over the NPUs, so that each NPU has a subset of the experts. In this scenario, when the gating function selects experts, tokens are only routed to the NPUs corresponding to the needed experts. When the experts perform their computations, another communication operation is needed to combine the experts' output tokens back to their corresponding DP instances.

\subsection{Disaggregating MoE and Attention}
xDeepServe offers two modes of deployment for MoE models. The first is an MoE-attention-collocated (\textbf{MA-collocated}) deployment, which resembles typical deployments where the attention modules, dense FFN, and MoE are hosted on the same device. 
As part of our initiative to develop a fully disaggregated LLM serving platform, we propose an MoE-attention-disaggregated (\textbf{MA-disaggregated}) deployment~\cite{stepfun2025,zhu2025}, which splits attention from MoE and hosts each module on separate devices. This provides some benefits such as independent scaling of each module, higher throughput, and better hardware utilization. The stateful attention modules handle the key-value (KV) cache and sequence scheduling, while the stateless MoEs execute in an infinite loop and perform forward computations whenever they receive any batches.  

Disaggregation introduces several challenges.
First, without microbatching, attention would issue a full batch to MoE and remaining idle until completion, causing inefficient hardware utilization. To avoid this, the batch is divided into multiple microbatches so attention and MoE can compute concurrently.
Second, traditional all-to-all collectives do not map directly to this architecture. Efficient group-to-group communication primitives are required to transfer tensors between attention and MoE modules.
Even with these optimizations, overall performance hinges on overlapping communication with computation to eliminate idle gaps~\cite{zhu2025}.

\subsection{XCCL}\label{sec:background:xccl}
Designed around the communications requirements of MoE models, XCCL~\cite{xiao2025} is our in-house communication library that offers collective communications for CloudMatrix384. XCCL utilizes the memory transfer units in the NPU to perform quick, low-latency operations, while direct memory access (DMA) is used for throughput-dependent operations. Specific to MoE models, XCCL offers \dispatch, which uses the gating scores to route tokens from attention to their top-k experts, and \combine, which aggregates expert outputs back to attention using the gating scores. In a disaggregated deployment, XCCL offers \lstinline|attention2expert| (\AtoE) and \lstinline|expert2attention| (\EtoA), which are analogous to \dispatch\ and \combine, but are designed to additionally handle communication asymmetry caused by differing numbers of attention and MoE NPUs.

\subsection{Graph Execution}
The typical execution mode in PyTorch, eager mode, will send single operations to the accelerator, which executes the operations as it receives them~\cite{torchgraphmode}. Eager mode supports dynamic tensor shapes and enables simple debugging, making it suitable for development purposes. However, it incurs a host-to-device communication cost for each kernel launch operation, which leads to poor NPU utilization for short-running fine-grained tensor operations. On the other hand, graph mode allows multiple operations to be fused together, known as operator fusion~\cite{torchgraphmode,xiao2025}. Using graph mode, multiple operations, even the entire model forward, can be completed with a single kernel launch, thus avoiding much of the host-to-device overhead. However, before such a computation graph can be used, it must be compiled by tracing through its operators, which is time consuming and hampers recovery times. To improve compilation times, many compilers offer the ability to perform a cached compile~\cite{cache_compile} using a cached version of a past compilation. We describe cached compilation further in \autoref{sec:graph}.
\section{\sysname\ Design}

\begin{figure}[t]
    \centering
    \begin{subfigure}[h]{0.45\linewidth}
    \centering
    \includegraphics[width=\linewidth]{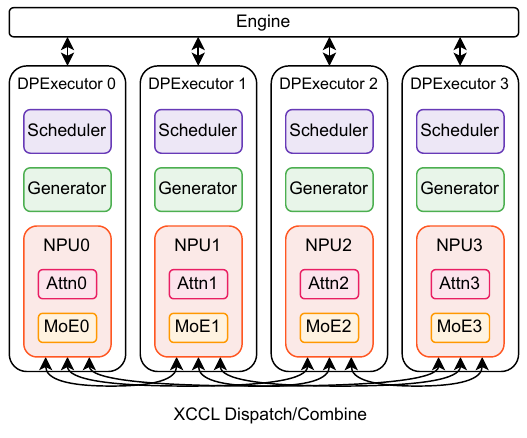}
    \caption{MA-collocated deployment style. Attention and MoE may be placed on the same processes and NPUs, with XCCL \dispatch\ and \combine\ performing communications.}
    \label{fig:collocated_flowserve}
    \end{subfigure}
    \hfill
    \begin{subfigure}[h]{0.45\linewidth}
    \centering
    \vspace{\baselineskip}
    \vspace{\baselineskip}
    \includegraphics[width=\linewidth]{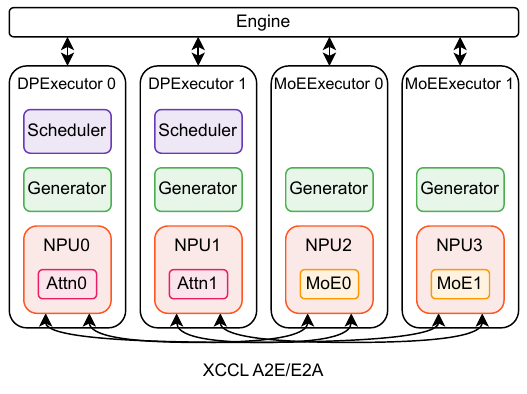}
    \caption{MA-disaggregated deployment style. Attention and MoE are placed on different processes and NPUs, with \AtoE\ and \EtoA\ operations performing communications between the attention and MoE groups.}
    \label{fig:ma_flowserve}
    \end{subfigure}
    \caption{Overview of a FlowServe inference instance.}
    \label{fig:flowserve}
\end{figure}

\begin{figure*}[t]
    \centering
    \includegraphics[width=0.9\linewidth]{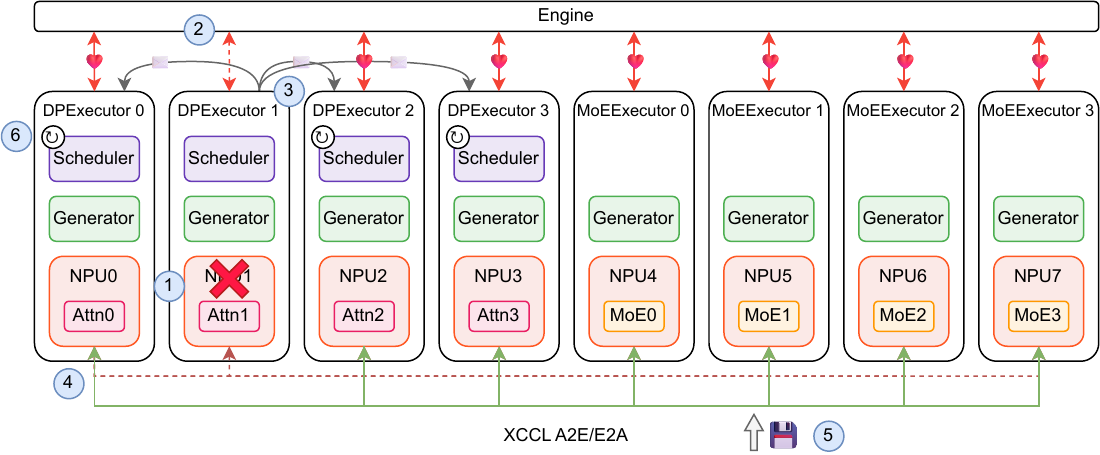}
    \caption{
    \sysname\ design under an attention failure scenario in a MA-disaggregated deployment.
    \circled{1}~NPU1 experiences a failure. 
    \circled{2}~The engine does not receive the heartbeat from DPExecutor 1, so it initiates recovery. 
    \circled{3}~Requests get migrated from DPExecutor 1 to other DPExecutors, and DPExecutor 1 gets terminated. 
    \circled{4}~The communications domain is destroyed and reinitialized without NPU1. 
    \circled{5}~The graph cache is loaded from disk and a cached compilation builds the computation graph. 
    \circled{6}~The block table is restored on all DPExecutors and inference can begin again. 
    }
    \label{fig:revivemoe_design}
\end{figure*}

Before discussing the design of \sysname, we provide some context and a high-level overview of FlowServe, which operates as a model serving instance under xDeepServe. \autoref{fig:flowserve} illustrates a simplified version of FlowServe's system design. First, we describe the MA-collocated case in \autoref{fig:collocated_flowserve}. A central engine governs the processes and performs global scheduling and dispatching of user requests to the executors. 
DPExecutors are separate processes that each host a local scheduler, generator, and are assigned one NPU, which hosts the model weights corresponding to the DP rank. The local scheduler controls which sequences proceed to generation and which sequences wait in each generation step. The generator calls the model forward steps and performs sampling. During inference, the ranks communicate with each other through XCCL's \dispatch\ and \combine\ function during the model forward. In the MA-disaggregated case in \autoref{fig:ma_flowserve}, we introduce the notion of a MoEExecutor, whose only role is to perform the forward step of the MoE and does not contain any attention-related modules like the KV cache. In this case, the attention weights and MoE weights are separated, and the communications leverage the \AtoE\ and \EtoA\ operation to perform necessary group-wise tensor transfers between ranks. 

We propose \sysname, a recovery system designed for FlowServe deployments. It targets the scenario of isolated, single NPU failures, and the failed NPU may never rejoin the cluster. Large-scale outages, such as network partitions or multiple nodes failures, are beyond the scope of \sysname. 
\autoref{fig:revivemoe_design} illustrates an example of the recovery process under a MA-disaggregated deployment. We propose a failure detection system (\autoref{sec:detection}) using heartbeats. Upon failure, we check whether these errors are in the failure scenarios covered by \sysname.
If they are, we initiate failure recovery and pause inference on all executors. Once recovery begins, user requests are migrated from the executor containing the failed hardware (\autoref{sec:migration}), with a partial recomputation strategy once they arrive in the new executor. The block table on attention ranks must be restored to a state prior to failure using a log-based recovery technique (\autoref{sec:block_table}). Even with some redundancy in DP deployments, MoE weights need to be carefully handled depending on the deployment configuration (\autoref{sec:weight}). Once weights are intact, the communication domain can be rebuilt to exclude the failed hardware (\autoref{sec:communication}), and the graph cache can be read from disk and compiled \autoref{sec:graph}). Finally, inference can then begin again without the failed hardware.

\subsection{Failure Detection}\label{sec:detection}
The Kubernetes Device Plugin interface~\cite{k8sdeviceplugin} enables management of specialized hardware like NPUs by allowing vendor-specific plugins to register devices as extended resources and handle allocation and monitoring via gRPC, without altering core Kubernetes code. Huawei's NPU device plugin \cite{huawei2025cce} implements this interface for Ascend NPU platforms, providing resource monitoring and fault reporting. It detects NPU health issues, reports them to kubelet, and logs detailed fault information such as the event ID, alarm time, severity, and error type. 
We define fault codes across six levels (L1–L6) based on severity, where L1 faults are benign and require no action, while L6 faults are critical and result in full isolation of the NPU. Our technique aims to cover all six of these levels. While the Huawei NPU Device Plugin reports these errors to Kubernetes node annotations, we enable proactive fault management by setting up a separate Ray actor within FlowServe to periodically monitor these annotations and trigger \sysname's failure recovery procedures.

\subsection{Sequence State Recovery}\label{sec:migration}

In order to recover sequences from failed attention ranks, we must migrate them from the failed hardware to other available hardware. In our migration strategy, we adopt a partial recomputation strategy by preserving the prompt and decoded tokens thus far. The sequences' KV caches are assumed to be missing due to failure. However, their corresponding tokens IDs are still available on the DPExecutor's CPU memory. This means that we can jointly preserve the prompt and any decoded token IDs by concatenating them into a new prompt and migrating them to an available attention rank. Upon service continuation, the new attention rank re-executes the prefill for the concatenated prompt but can skip decoding steps that were previously completed

As each decoding step proceeds, executors may receive the global signal to stop and exit upon a detected failure. We adopt step-level recovery because layer-level checkpoints can leave inconsistent KV states across layers, leading to corrupted cache reuse. We revert back to the start of the generation step by restoring the block table (\autoref{sec:block_table}) and recompute the next token while discarding any newly generated caches. This would perform the full model forward again when service continues. 

\subsection{Block Table Recovery}\label{sec:block_table}
During decoding, each sequence in the batch appends entries to its KV cache blocks. When a block is full, a new one is allocated on the next generation step using the block manager and block table. In the case where a failure occurs during a generation step, we must undo any block operations to keep the block table consistent and return it to the start of the generation step.
We achieve this by logging block operations similarly to log-based recovery in database management systems~\cite{mohan1992}. At the start of the current generation step, we clear the log and start a new one, as the previous step fully completed. Every time a block operation occurs, we append the operation to the log. Should any failure occur, we undo all operations in the current log, which brings us back to the start of the step. For example, undoing an allocation involves decrementing the block’s reference count or deleting it if unreferenced.

\subsection{Weight Integrity}\label{sec:weight}

\begin{figure}[]
    \centering
    \includegraphics[width=0.5\linewidth]{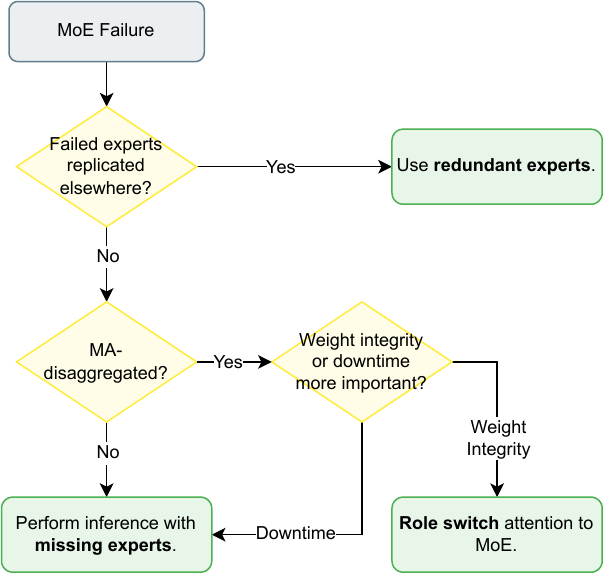}
    \caption{Flowchart for deciding action when a failure involves MoE weights.}
    \label{fig:moe_flowchart}
\end{figure}

As hardware is lost, so are any corresponding model weights. For attention weights, DP ensures that they are replicated across all DP groups. Moreover, we run attention TP=1 for better decoding throughput and compatibility with XCCL, which additionally means attention failures do not leave unusable attention weight shards in the system. For MoE weights, redundancy and weight integrity depend on the deployment configuration, and we offer three options to handle them: (1) redundant experts, (2) role switching, and (3) missing experts. \autoref{fig:moe_flowchart} illustrates a flowchart of these outcomes when an MoE failure occurs. 

\textbf{Redundant experts.} The most convenient solution is to use redundant experts, which we already use for load balancing purposes~\cite{dsv32025}. As their name suggests, they not only provide latency and throughput benefits, but also grant redundancy by replicating experts. Given that each expert on the failed NPU is replicated elsewhere at least one additional time, we can ensure that all model weights are still present in the system in the event of a single hardware failure. This procedure simply requires removing the failed experts from the logical-to-physical mapping that determines which NPUs each replica resides on. 

\textbf{Role switch.} For MA-disaggregated setups without enough redundant experts, an attention rank (already replicated) can be switched to replace the failed MoE rank. This involves selecting a DPExecutor to be switched to a MoEExecutor. The role switch process begins by migrating its requests to other available DPExecutors, then removing the KV caches, the local scheduler, and removing all attention weights. The logical ranks of all devices in the domain are also rearranged so that this new MoEExecutor takes the place of the failed MoEExecutor. Without MoE redundancy in this scenario, the only copies of these weights are lost due to failure. New MoE weights must be loaded from disk to the new MoEExecutor to replace the lost weights, which makes this method the most costly in terms of downtime. 

\textbf{Missing experts.} Without redundancy, an alternative is to allow some MoE weights to be lost. With sufficiently large EP, the number of experts lost during a single NPU failure may be small enough such that model accuracy is not meaningfully affected. We verify our accuracy claims in \autoref{sec:eval:accuracy}, and find that on a DeepSeek V3 model, $\frac{1}{32}$ of the experts may be lost (single MoE NPU failure in EP32) with minimal accuracy penalty. Under this scenario, a mask is added to the gating function to mask out the logits corresponding to the missing experts. This will ensure that the top-k set of experts chosen do not contain the missing experts and that the next-best experts are used in their place.

Although we had only discussed MoE weights for the FFN so far, some models such as DeepSeek V3 and Kimi K2 use dense FFNs in their first 1-3 layers instead of MoE. In our deployments, these dense FFNs are run in TP=4 and replicated over multiple FFN TP groups. If any dense FFN weight shard in the FFN TP group is lost due to failure and is not recovered, it is considered a compromised FFN TP group. We change the routing such that the attention modules evenly rebalance their outgoing tokens over the healthy dense FFN TP groups instead.

\subsection{Recreating Communications}\label{sec:communication}
In recreating communications domains, we treat the failed NPU as inaccessible, meaning that it physically still exists in the system, but we cannot perform any operations with it. This comes into play when we assign ranks in the various communication groups. In PyTorch process groups using GLOO~\cite{gloo} or HCCL~\cite{hccl}, we keep the default world group intact but reassign subgroups such as the DP and EP groups so that they do not contain the failed rank. In XCCL, we must fully destroy and recreate the domain. This involves destroying the trampoline~\cite{xiao2025} domain between experts for MA-disaggregated deployments, then a universal step of destroying the communication domain between attention and experts. To recreate XCCL domains, we must assign new logical ranks to compact the communication domain. For example, if NPU A with logical rank $\ell_A$ fails, it leaves a gap in rank assignments. We reassign NPU B with logical rank $\ell_B = \ell_A + 1$ to $\ell_A$ and decrement subsequent ranks to close the gap. In the role switching case, switched NPU C with logical rank $\ell_C$ takes the logical rank $\ell_A$ of failed NPU A. Then we fill in any gaps according to the previous procedure.
Using this new assignment, we can create the XCCL attention-expert domain.

\subsection{Graph Mode}\label{sec:graph}
Graph mode improves decoding throughput and latency by minimizing kernel launch overhead~\cite{xiao2025}. However, it requires a lengthy compilation before it can be used. As the amount of hardware we have is now reduced by 1 due to the failure, we must now recompile a new graph as the previous graph was compiled for the old deployment size. In our findings, compiling the full DeepSeek V3 model took 12.9 minutes, which even dominates weight loading times. However, these times can be reduced through a procedure known as cached compilation~\cite{cache_compile}. Typical graph compilation requires compiling Dynamo~\cite{dynamo} and the Ascend IR graph, which are the most time-consuming steps. 
The results of these two steps can be saved to disk as a cache so that they do not need to be recompiled. Future compilations with the same system configuration can reuse this cache to accelerate compilation.
We leverage cached compilation into our recovery procedure by precompiling a graph cache under a failure scenario, so that when a failure occurs, we have a cache ready to perform a cached compile.

\begin{figure*}[t]
    \centering
    \includegraphics[width=\linewidth]{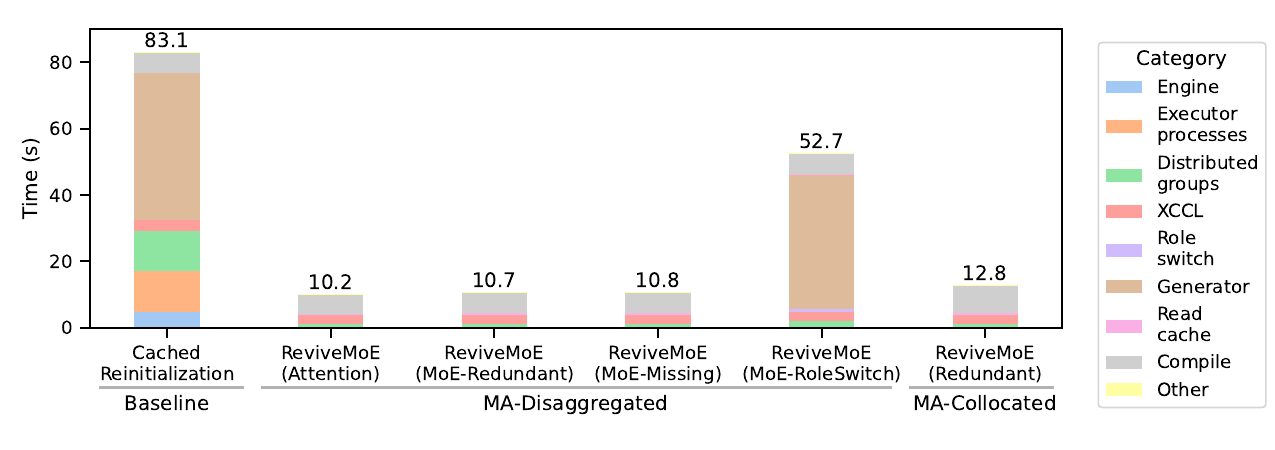}
    \caption{Recovery times for various \sysname\ scenarios. In the MA-disaggregated scenario, the brackets indicate the module failed and MoE recovery technique, if applicable. \autoref{tab:labels} further explains each timing category.}
    \label{fig:recovery_time}
\end{figure*}

\begin{table*}[t]
    \centering
    \caption{Explanation of timing categories in evaluation.}
    \begin{tabular}{p{1.5in}p{4.5in}}
    \hline
      Category   &  Description\\
      \hline
      Engine & Time to initialize the engine.\\
      Executor Processes & Time to launch all executor processes, run their constructors, and allocate them to Ray resources.\\
      Distributed Groups & Time to setup the torch distributed groups using HCCL and GLOO.\\
      XCCL & Time to form the XCCL communication domain.\\
      Role Switch & Time to role switch a DPExecutor to MoEExecutor.\\
      Generator & Time to initialize the generator, including model parameters, weight loading, and warmup operations for the KV cache.\\
      Read Cache & Time to load the cached graph from disk.\\
      Compile   & Time to perform a cached compile for the computation graph.\\
      Other & Any small overhead that does not individually exceed 100 ms, such as scheduler initialization, task cancellations, and migration.\\
      \hline
    \end{tabular}
    \label{tab:labels}
\end{table*}

\section{Evaluation}\label{sec:eval}
To evaluate \sysname\ on its recovery capabilities, we look at its recovery time in \autoref{sec:eval:recovery}, which is representative of the downtime a system experiences upon failure. In \autoref{sec:eval:accuracy}, we perform an experiment to justify our claims that having missing experts is a valid solution provided sufficient EP.

In all experiments, we deploy the DeepSeek V3 model~\cite{dsv32025} on the CloudMatrix384~\cite{zuo2025}. We use a subset of the available NPUs composed of 80 Huawei Ascend 64 GB chips. Software-wise, we use NPU driver version 25.2.1 with CANN 8.2.1 on Torch NPU 2.1.0.

\subsection{Recovery Time}\label{sec:eval:recovery}

To show the effectiveness of our method, we simulate the failure of a single card in the system and initiate recovery. We compare \sysname\ with the baseline technique of reinitializing the system using a compilation cache. We measure the cached reinitialization time as solely the time to initialize FlowServe, meaning that the Docker containers and Ray are assumed to be available and their times are not included, but FlowServe must relaunch the engine and all executor processes. It must then perform the corresponding weight loads, communications operations, and cached graph compilations. In our recovery technique, we subdivide it to evaluate cases in which an attention rank fails cases in which a MoE rank fails. In the case where a MoE rank fails, it requires either role switching an attention rank to MoE, leveraging redundant experts, or allowing for lost experts.  The figure of merit here is the recovery time, which represents the downtime the system experiences upon failure and recovery. 

\autoref{fig:recovery_time} summarizes our results with \autoref{tab:labels} explaining our timing categorizations. Firstly, looking at the MA-disaggregated case, our technique outperforms the baseline in every case, saving up to 87.8\% in recovery time. Even in the case where there is a role switch and the weights must be loaded, we still save 36.6\% because the engine and executor processes do not need to be relaunched. When role switching is not required, the timings are nearly identical, as the steps are generally the same, with attention recovery requiring a migration step and MoE recovery requiring an update to their gating mechanisms, which all occur in under 50 ms. In the case where role switching is required, the time is dominated by the weight loading required in the generator, as attention weights must be discarded and replaced with MoE weights, which takes 40.6~s. The most significant reduction, which is not portrayed in the figures, is from avoiding a full graph compilation. Since we precompile the graph caches for a failure case, this reduces the compilation overheads to 6~s. However, if we were to do a full compilation without the caches, it would take 12.9 minutes, which highlights the necessity of quick compilations when switching system configurations. The MA-collocated case behaves largely similar to the MA-disaggregated redundant expert case, as its steps are identical. The difference is mainly in compilation complexity (8~s vs. 6~s) due to its joint attention–MoE computation.

\subsection{Model Accuracy with Lost Experts}\label{sec:eval:accuracy}

\begin{figure}[t]
    \centering
    \includegraphics[width=0.5\linewidth]{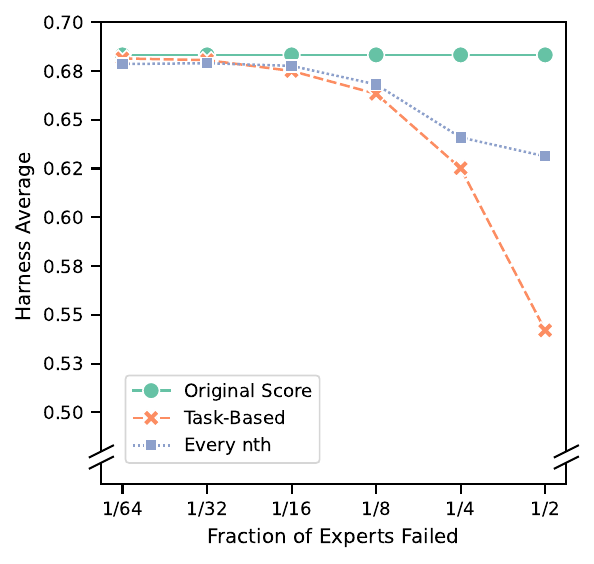}
    \caption{DeepSeek V3 Harness average as experts are lost.}
    \label{fig:expert_failure}
\end{figure}

\begin{table*}
\caption{DeepSeek V3 Harness score on each task as experts are lost.}
\centering
\fontsize{9}{10}\selectfont
\setlength{\tabcolsep}{5.5pt}
    \begin{tabular}{lccccccccccccc}

\toprule
 & \multicolumn{1}{c}{Base} & \multicolumn{6}{c}{Task-Based} &  \multicolumn{6}{c}{Every nth}\\ 
\cmidrule(lr){2-2} \cmidrule(lr){3-8}  \cmidrule(lr){9-14}
Fraction Lost $r$ & - & 1/64 & 1/32 & 1/16 & 1/8 & 1/4 & 1/2 & 1/64 & 1/32 & 1/16 & 1/8 & 1/4 & 1/2 \\
\midrule
ARC Challenge & 0.647 & 0.664 & 0.656 & 0.660 & 0.652 & 0.612 & 0.524 & 0.650 & 0.648 & 0.648 & 0.630 & 0.570 & 0.548 \\
ARC Easy & 0.862 & 0.877 & 0.875 & 0.867 & 0.865 & 0.840 & 0.782 & 0.861 & 0.859 & 0.857 & 0.851 & 0.819 & 0.806 \\
WinoGrande & 0.811 & 0.807 & 0.801 & 0.797 & 0.794 & 0.781 & 0.734 & 0.806 & 0.809 & 0.809 & 0.793 & 0.787 & 0.771 \\
HellaSwag & 0.868 & 0.865 & 0.864 & 0.861 & 0.850 & 0.830 & 0.741 & 0.860 & 0.861 & 0.861 & 0.855 & 0.832 & 0.815 \\
PIQA & 0.860 & 0.849 & 0.850 & 0.838 & 0.828 & 0.811 & 0.751 & 0.848 & 0.847 & 0.846 & 0.845 & 0.828 & 0.827 \\
RACE & 0.447 & 0.447 & 0.444 & 0.461 & 0.460 & 0.451 & 0.433 & 0.443 & 0.447 & 0.440 & 0.438 & 0.435 & 0.429 \\
TruthfulQA MC1 & 0.348 & 0.324 & 0.334 & 0.333 & 0.335 & 0.343 & 0.299 & 0.344 & 0.346 & 0.339 & 0.356 & 0.333 & 0.326 \\
TruthfulQA MC2 & 0.504 & 0.480 & 0.496 & 0.491 & 0.506 & 0.507 & 0.461 & 0.497 & 0.495 & 0.496 & 0.500 & 0.479 & 0.476 \\
GSM8k & 0.641 & 0.658 & 0.650 & 0.616 & 0.540 & 0.318 & 0.111 & 0.633 & 0.635 & 0.637 & 0.589 & 0.534 & 0.523 \\
MMLU & 0.845 & 0.842 & 0.835 & 0.825 & 0.803 & 0.760 & 0.584 & 0.843 & 0.843 & 0.843 & 0.823 & 0.791 & 0.791 \\
\midrule
Average & 0.683 & 0.681 & 0.681 & 0.675 & 0.663 & 0.625 & 0.542 & 0.679 & 0.679 & 0.678 & 0.668 & 0.641 & 0.631 \\
\bottomrule
\end{tabular}
\label{tab:expert_failure}
\end{table*}

To verify our claim that few lost experts minimally affect accuracy in \autoref{sec:weight}, we perform an evaluation by selectively failing a fraction of the experts $r\in \left\{\frac{1}{64},\frac{1}{32},\frac{1}{16},\frac{1}{8},\frac{1}{4},\frac{1}{2}\right\}$ and evaluate accuracy on various tasks. In this experiment, we define a failed expert $i$ such that when $i$ is chosen, it is treated as failed across all MoE layers. We avoid the model selecting $i$ by masking its routing logit to negative infinity immediately before the top-k selection. We select failed experts under two scenarios. The first scenario, \textbf{task-based}, emulates a worst-case scenario by failing the most selected experts per task. We first run DeepSeek V3 on each calibration dataset, count the activated experts per layer, aggregate the counts across layers to obtain a global ranking, then fail fraction $r$ of the most frequently activated experts and rerun the corresponding evaluation. The second scenario, \textbf{every nth}, emulates a more uniform scenario by failing experts at a step-size to target failing fraction $r$ of experts. For example, every even-indexed expert fails across all layers when $r=\frac{1}{2}$. 
We perform our evaluation on DeepSeek V3 using the Language Model Evaluation Harness~\cite{eval-harness} on the following tasks: ARC Challenge/Easy~\cite{clark2018}, WinoGrande~\cite{sakaguchi2019}, HellaSwag~\cite{zellers2019}, PIQA~\cite{bisk2019}, RACE~\cite{lai2017}, TruthfulQA MC1/MC2~\cite{lin2022}, GSM8k~\cite{cobbe2021}, and MMLU~\cite{hendrycks2021}.

\autoref{tab:expert_failure} and \autoref{fig:expert_failure} summarize our results. Each row of \autoref{tab:expert_failure} represents the score for each task. The columns indicate the expert failure scenario, where base is the default scenario with no failed experts, and the other groups are the scenarios that fail fraction $r$ of experts. \autoref{fig:expert_failure} is a plot of the arithmetic mean over all of these tasks for each scenario. We find that up to $\frac{1}{32}$ of experts can be lost with minimal effect to accuracy. To relate these results to our single NPU failure case, consider that the fraction of experts failed is the inverse of the deployment EP, e.g., $\frac{1}{64}$ experts lost corresponds to a single MoE failure in EP64. We see that in both the task based and every nth case, up to $\frac{1}{32}$ experts can be lost with minimal effect on accuracy. In other words, in a single failure case, we can lose a single MoE NPU in EP32 or higher with minimal accuracy loss. 

\subsection{Necessity of Role Switching}
As we saw in \autoref{sec:eval:accuracy} and \autoref{sec:eval:recovery}, DeepSeek V3 is robust to expert losses up to $\frac{1}{32}$ of the experts, while redundant experts do not compromise recovery time nor accuracy. This may raise the question as to why we would want role switching. 
Role switching becomes a factor in two scenarios: (1) when EP $<32$, where lost experts have non-negligible impact, or (2) when the final copy of a redundant expert is lost. In practice, redundant experts are typically selected based on usage frequency~\cite{dsv32025} rather than fault tolerance, so low-use experts may not be replicated. As a result, even with redundancy, the loss of the last copy of an expert can necessitate a role switch.
These techniques are not mutually exclusive and can be combined within the recovery procedure. For example, a role switch can begin in the background while the system continues inference using the current (possibly incomplete) expert set. This approach enables both rapid recovery and eventual restoration of full weight integrity.

\section{Related Work}
We provide an overview of some literature that describe the topics of reliability and fault tolerance for LLMs.
\subsection{Large-Scale Cluster Reliability}
Several works have analyzed errors that occur in large-scale ML clusters. 
IBM~\cite{gershon2025} discuss their Vela and Blue Vela AI infrastructure and introduce a taxonomy for infrastructure failure types. They additionally state that they encountered hardware failure rates of 2-5\% per month, with many requiring vendor repair, which highlights the need for recovery techniques that do not rely on hardware to return quickly.
Meta~\cite{kokolis2025} released a similar report where they introduce a taxonomy of failures in research clusters at Meta, use their historical data to estimate failure times, and propose preventative techniques such as identifying and repairing vulnerable nodes. Microsoft~\cite{gao2023} perform a similar analysis, and investigate root causes of quality issues in ML clusters. Duan et al.~\cite{duan2024} overview LLM training reports from various companies and corroborate many cases of high failure rates, with one notable instance where 56\% of the actual training time was spent dealing with failures when Meta was training their OPT 175B model~\cite{zhang2022}.
Our work is complementary to their analysis, where we focus primarily on the recovery process once an inevitable failure occurs instead of preventing failures.

\subsection{ML Inference Recovery}
Two prior studies address recovery in an inference setting. Liu et al.~\cite{liu2025} introduce an MA-disaggregated design with fault tolerance in mind. Their failure recovery does not need to reinitialize the communication domain, as they leverage peer-to-peer operations for their communications. However, this is unsuitable for performance on NPUs, and we must use collective operations. In addition, their work solely relies on redundant experts as the recovery option for MoE on MA-disaggregated designs, whereas we offer 3 options that also include MA-collocated designs. Strati et al.~\cite{strati2024} propose a recovery technique for pipeline parallel failures. They use KV cache replication to restore missing KV caches when a pipeline stage recovers. Their approach assumes that the failed hardware quickly returns, whereas we do not make such an assumption. In addition, their work focuses on pipeline parallelism, while our focus is on deployments using DP, TP, and EP.

\subsection{ML Training Recovery}

Checkpointing is frequently used in ML training to recover from failures by periodically saving model weights and restoring them upon failure. Much of the recovery time is then spent restoring the checkpoint, then replaying the training process to get back to the state prior to failure. Many works in this domain describe better checkpointing techniques \cite{mohan2021, duan2024, gupta2024}.
FlashRecovery~\cite{zhang2025} focuses on training recovery speed and proposes a fast training recovery method that does not need checkpointing. They propose a similar philosophy in that they avoid full reinitialization, and instead try to isolate the failure then work around it. They also implement a heartbeat-based failure detection mechanism, isolated task termination, and live communication reinitialization. However, their work contrasts with ours in that they replace failed hardware with new hardware, while we operate under a setting where failed hardware cannot easily be replaced, and we must continue service while missing hardware. Overall, while some lessons from training recovery may be applicable to our problem, 
\sysname\ focuses on recovery during inference, which typically has tighter requirements from customer-facing service demands.
\section{Conclusion}
We have presented \sysname, which allows for quick recovery in model-as-a-service inference settings without costly instance reinitialization. We discuss several techniques that act together to provide quick recovery: a failure detection method using heartbeats, a log-based recovery system for the block table, weight integrity preservation techniques, communication reinitialization steps, and cached execution graph compilation. 

We identify some limitations with our current design and envision future avenues. 
Currently, we only handle failure detection in the case where there is an obvious failure. Slowdowns or power issues are not as obvious but should be handled, as even a single slow device can cause significant delays in the overall system due to communication synchronization in MoE models. In an alternative direction, larger-scale failures are not yet handled by \sysname. An extension to solve this is that redundant expert placement would need to balance both performance and fault tolerance to handle node-level failures. Other large-scale failures, such as network partitions, are difficult to deal with under tight SLO constraints, even in typical distributed systems. Future avenues can aim to address these issues of hardware slowdowns and larger-scale failures.

\bibliography{bib}

\end{document}